\documentclass[11pt,showpacs,preprint,preprintnumbers,nofootinbib,superscriptaddress,amsmath,amssymb]{revtex4}

\usepackage{graphicx}
\usepackage{dcolumn}
\usepackage{bm}
\usepackage{amssymb}
\usepackage{amsmath}
\usepackage{epsfig}    
\usepackage{color}
\usepackage{slashed}

\begin{document} 

\title{The 750 GeV diphoton excess and its explanation\\
 in 2-Higgs Doublet Models with a real inert scalar multiplet}

%
\author{Stefano Moretti}
\email{S.Moretti@soton.ac.uk}
\affiliation{School of Physics and Astronomy, University of Southampton, Southampton, SO17 1BJ, United Kingdom}
\author{Kei Yagyu}
\email{K.Yagyu@soton.ac.uk}
\affiliation{School of Physics and Astronomy, University of Southampton, Southampton, SO17 1BJ, United Kingdom}

\begin{abstract}

We discuss a possible explanation of  the recently observed diphoton excess at around 750 GeV as seen by the ATLAS and CMS experiments at the Large Hadron Collider. 
We calculate the cross section of the diphoton signature 
in 2-Higgs Doublet Models  with the addition of a real isospin scalar multiplet without a vacuum expectation value, where 
a neutral component of such a representation can be a dark matter candidate. 
We find that the branching fraction of an additional CP-even Higgs boson $H$ from the doublet fields 
into the diphoton mode  
can be significantly enhanced,  by up to a factor of $10^3$, with respect to the case of the simple 2-Higgs Doublet Model. 
Such a sizable enhancement is realized due to multi-charged inert particle loops entering the $H\to \gamma\gamma$ decay 
mode.  
Through this enhancement, we obtain a suitable cross section of the $gg\to H \to \gamma\gamma$ process to explain the data with the fixed 
mass of $H$ being 750 GeV. 

\end{abstract}
\maketitle

\section{Introduction}

Recently, there was a quite exciting report about a new excess at around 750 GeV in the diphoton invariant mass distribution 
at the 13 TeV run of the CERN Large Hadron Collider (LHC)
from both ATLAS and CMS groups. 
The excess was observed with a local significance around 3.6$\sigma$ at ATLAS~\cite{750GeV-ATLAS} using 3.2 fb$^{-1}$ data and around 2.6$\sigma$ at CMS~\cite{750GeV-CMS}  
using 2.6 fb$^{-1}$ data. 
Most instinctively, this excess can be explained in such a way that there is a new neutral spin-0 resonant particle\footnote{Although there could exist in principle a spin 2 particle as an alternative explanation, 
it seems difficult to construct a phenomenologically 
reasonable model to include such a state.  
In contrast, a spin 1 resonant state cannot directly decay into the diphoton final state~\cite{Moretti:2014rka}, 
but it can decay into the triphoton state via a scalar boson inter mediation~\cite{spin1}. 
} with a mass of about 750 GeV which decays into two photons. If this conjecture is correct and the excess is confirmed by further data, this will represent direct evidence for the existence of new physics  Beyond the Standard Model (BSM).

So far, there appeared a number of papers to describe the excess since it was reported.  
For example, in Refs.~\cite{effective}, the excess was explained via a scalar boson resonance by introducing an effective Lagrangian with dimension five operators.  
It was also explained in a model with an extra isospin singlet scalar field~\cite{singlet-0,singlet,hsm-thdm}, that with an extra doublet scalar field including 
the minimal supersymmetric Standard Model (SM)~\cite{750GeV-THDM,thdm,hsm-thdm} and that with an extra triplet scalar field~\cite{triplet,331}. 
A connection between the Dark Matter (DM) physics and the diphoton excess has been discussed in Ref.~\cite{dm-750}. 

We discuss the possibility to explain this excess in 2-Higgs Doublet Models (2HDMs) supplemented by a suitable	
 additional scalar representation. The latter addition to a standard 2HDM spectrum is needed, 
as Ref.~\cite{750GeV-THDM} has already pointed out that the gluon fusion process $gg \to H/A \to \gamma\gamma$ within 2HDMs, 
where $H$ and $A$ are the additional CP-even and CP-odd Higgs bosons, respectively, is not sufficient to produce the event yield required, essentially because the 
the branching fraction of $H/A\to \gamma\gamma$ is typically of ${\cal O}(10^{-3})$ smaller than the necessary value.

In fact, there are two ways to reproduce the required cross section, of order 10 fb: to have an enhancement in (i) the gluon fusion cross section and/or  
(ii) in the diphoton branching fraction. 
In this paper, we consider the second possibility, by introducing an additional real inert scalar multiplet without a non-zero Vacuum 
Expectation Value (VEV) in 2HDMs with a softly-broken $Z_2$ symmetry. 
Such an inert scalar multiplet is for example motivated in the ``minimal DM scenario'' discussed in Refs.~\cite{minimal-DM,Garcia}, where 
a neutral component of the multiplet can be a viable DM candidate. 
Thanks to the introduction of the inert multiplet, the branching fraction of $H\to \gamma\gamma$ is significantly enhanced, indeed by a factor of $10^{2}$-$10^{3}$ depending 
on the isospin of the inert multiplet. 

The plan of the paper is as follows. In Sec.~II, we define our model. In Sec.~III, 
we calculate the gluon fusion production cross section and the decay branching ratio into the diphoton state
of the additional neutral Higgs bosons. 
By combining these, we show the cross section of the diphoton process. 
The cutoff scale of our model is estimated by using one-loop renormalization group equations (RGEs). 
We then conclude in Sec.~IV.
In Appendix, we present the set of beta functions for all the relevant coupling constants at the one-loop level. 

\section{The Model} 

\begin{table}[t]
\begin{center}
\begin{tabular}{c||c|c|c|c||c|c|c}
\hline\hline &  $u_R^{}$ & $d_R^{}$ & $e_R^{}$ &
 $Q_L$, $L_L$ &$\xi_u$ & $\xi_d$ & $\xi_e$\\  \hline
Type-I   & $-$ & $-$ & $-$ & $+$ & $\cot\beta$ & $\cot\beta$ & $\cot\beta$\\
Type-II  & $-$ & $+$ & $+$ & $+$ & $\cot\beta$ & $-\tan\beta$ & $-\tan\beta$\\
Type-X   & $-$ & $-$ & $+$ & $+$ & $\cot\beta$ & $\cot\beta$ & $-\tan\beta$\\
Type-Y   & $-$ & $+$ & $-$ & $+$ & $\cot\beta$ & $-\tan\beta$ & $\cot\beta$\\
\hline\hline
\end{tabular} 
\end{center}
\caption{The $Z_2$ charge assignments for the SM fermion fields.  
The $\xi_f$ factors are given for each 2HDM type.} \label{xif}
\end{table}

We consider an extension of 2HDMs where 
the scalar sector is composed of two isospin doublets $\Phi_1$ and $\Phi_2$ with  hypercharge $Y=1/2$ 
and a real ($Y=0$) inert scalar multiplet $\chi$ with  isospin $T$ with a null VEV.
In particular, we consider the case for $T=2,~3$ and 4, or equivalently $\chi$ is assumed to be an isospin quintuplet, septet and nonet, respectively.   
We note that the maximal allowed value of $T$ has been obtained to be 4 from  perturbative unitarity arguments\footnote{In Ref.~\cite{Tsumura}, 
the scale dependence on dimensionless scalar couplings has been calculated in a model with one $SU(2)_L$ doublet Higgs field and a higher isospin 
multiplet using one-loop renormalization group equations. 
It has been clarified that if we have a $T\geq 3/2$ multiplet, the Landau Pole appears below the Planck scale. }~\cite{Logan}. 

The active sector involving $\Phi_1$ and $\Phi_2$ which have non-zero VEVs 
is similar to that of 2HDMs, so that 
there appears flavor changing neutral currents (FCNCs) at the tree level. 
In order to forbid such phenomena, we impose a softly-broken discrete $Z_2$ symmetry~\cite{GW} under which the scalar fields transform as 
$(\Phi_1,\Phi_2,\chi)\to (+\Phi_1,-\Phi_2,+\chi)$. 
Under the $Z_2$ symmetry, there are four types of Yukawa interactions~\cite{Barger,Grossman} 
and they are called Type-I, Type-II, Type-X and Type-Y~\cite{typeX} depending on 
the $Z_2$ charge assignment for right-handed fermions. 
In Table~\ref{xif}, the $Z_2$ charge assignment for all fermion fields is shown in each types of Yukawa interactions. 

The most general scalar potential under the $SU(2)_L\times U(1)_Y\times Z_2$ invariance with  CP-conservation is given by~\cite{inert-potential} 
\begin{align}
V(\Phi_1,\Phi_2,\chi) &=  \mu_1^2 \Phi_1^\dagger \Phi_1 + \mu_2^2 \Phi_2^\dagger \Phi_2 - \mu_3^2 (\Phi_1^\dagger \Phi_2 + \text{h.c.}) + \mu_\chi^2 \chi^\dagger \chi \notag\\
 & +\frac{1}{2}\lambda_1 (\Phi_1^\dagger \Phi_1)^2 +\frac{1}{2}\lambda_2 (\Phi_2^\dagger \Phi_2)^2  +\lambda_3(\Phi_1^\dagger \Phi_1)(\Phi_2^\dagger \Phi_2)
+\lambda_4|\Phi_1^\dagger \Phi_2|^2 +\frac{1}{2}\lambda_5[(\Phi_1^\dagger \Phi_2)^2 + \text{h.c.} ] \notag\\
& +\lambda_\chi (\chi^\dagger \chi)^2  +\lambda_\chi' (\chi^\dagger T^aT^b\chi)^2  
+ \rho_1(\Phi_1^\dagger\Phi_1)(\chi^\dagger \chi)
+ \rho_2(\Phi_2^\dagger\Phi_2)(\chi^\dagger \chi),     \label{pot}
\end{align}
where $T^a$ ($a=1,2,3$) are the $(2T+1)$ dimensional representation of the $SU(2)$ generator. 
We note that the operator $(\chi^\dagger T^a\chi)$  is identically zero~\cite{inert-potential}.
We also note that for the case of $T=2$, the $\lambda_\chi'$ term is written by the $\lambda_\chi$ term, so that 
we can take $\lambda_\chi'=0$. 
On the other hand for the $T=3$ and 4 case, the $\lambda_\chi'$ term gives the independent combination of the component fields of $\chi$. 

There are five physical scalar states from $\Phi_1$ and $\Phi_2$ as in  2HDMs, i.e., 
one pair of singly-charged Higgs bosons $H^\pm$, a CP-odd Higgs boson $A$ and two CP-even Higgs bosons $h$ and $H$
(with $m_h < m_H^{}$), where 
$h$ is assumed to be the discovered Higgs boson with a mass of about 125 GeV. 
The ratio of the two doublet VEVs $v_1$ and $v_2$ is defined as $\tan\beta = v_2/v_1$ and 
the two VEVs satisfy $v_1^2+v_2^2 = v^2 = (\sqrt{2}G_F)^{-1/2}\simeq (246$ GeV$)^2$. 
The mass formulae for these Higgs bosons are the same as those in 2HDMs (see e.g., Ref.~\cite{Chiang-Yagyu} for the explicit expressions).  

The squared mass of $\chi$ is given by 
\begin{align}
m_{\chi}^2=\mu_\chi^2+\frac{v^2}{2}(\rho_1 \cos^2\beta +\rho_2 \sin^2\beta), 
\end{align}
where all the masses of the component scalar fields in $\chi$ are degenerate at the tree level. 
A non-zero mass splitting can be generated via  one-loop corrections whose amount is typically only ${\cal O}(100)$ MeV as shown in Ref.~\cite{minimal-DM}. 
In the following discussion, we thus neglect such a small mass difference. 

\section{Diphoton excess}

We now discuss how we can reproduce the diphoton excess at around 750 GeV at the LHC in our  BSM scenario. 
Herein, the additional neutral Higgs bosons $H$ and $A$ can contribute to this excess via the gluon fusion process, 
i.e., $gg\to H/A \to \gamma\gamma$, by taking the mass of $H$ and $A$ to be 750 GeV. 
The production cross section of $gg\to H/A$ is calculated as follows 
\begin{align}
\sigma(gg\to H/A) = \sigma (gg \to h_\text{SM})\times \frac{\Gamma(H/A\to gg)}{\Gamma(h_{\text{SM}}\to gg)}, 
\end{align} 
where $h_{\text{SM}}$ is the Higgs boson in the SM with a mass of 750 GeV. 
The SM cross section $\sigma (gg \to h_\text{SM})$ with $m_{h_{\rm SM}}=$ 750 GeV 
at the collision energy of 13 TeV is given by about 736 fb, as quoted from 
the LHC Higgs Cross Section Working Group page~\cite{SM-cross}. 
The decay rate is given by 
\begin{align}
\Gamma({\cal H}\to gg)&=\frac{\sqrt{2}G_F\alpha_s^2 m_{\cal H}^3}{128\pi^3 }\left|\sum_f \xi_{\cal H}^fF_{1/2}^{\cal H}(m_f)\right|^2,\quad {\cal H} = H,~h, \\
\Gamma(A\to gg)       &=\frac{\sqrt{2}G_F\alpha_{s}^2m_A^3}{128\pi^3 }\left|\sum_f 2T_f^3 \xi_f F_{1/2}^A(m_f) \right|^2, 
\end{align}
where $T_f^3$ is the third component of the isospin of $f$, i.e., $T_f^3=+1/2\,(-1/2)$ for $f=u\,(d,e)$, and 
the $\xi_{{\cal H}}^f$ factors are expressed in terms of the $\xi_f$ expressions given in Tab.~\ref{xif} by 
\begin{align}
\xi_h^f &= \sin(\beta-\alpha)+\xi_f \cos(\beta-\alpha), \\
\xi_H^f &= \cos(\beta-\alpha)-\xi_f \sin(\beta-\alpha). 
\end{align}
We note that the decay rate for $h_{\text{SM}}$ is obtained from $\Gamma(h\to gg)$ by taking $\sin(\beta-\alpha)=1$. 

The loop functions for the CP-even Higgs bosons $F_{1/2}^{\cal H}(m_f)$ and the CP-odd Higgs boson $F_{1/2}^A(m_f)$ are given by 
\begin{align}
F_{1/2}^{\cal H}(m_f)  & = -\frac{4m_f^2}{m_{\cal H}^2}\left[2-m_{\cal H}^2\left(1-\frac{4m_f^2}{m_{\cal H}^2}\right)C_0(0,0,m_{\cal H}^2,m_f,m_f,m_f)\right], \\
F_{1/2}^A(m_f)  & =-4m_f^2C_0(0,0,m_A^2,m_f,m_f,m_f), 
\end{align}
where $C_0$ is the usual three-point Passarino-Veltman function~\cite{PV}.  

\begin{figure}[t]
\begin{center}
\includegraphics[width=80mm]{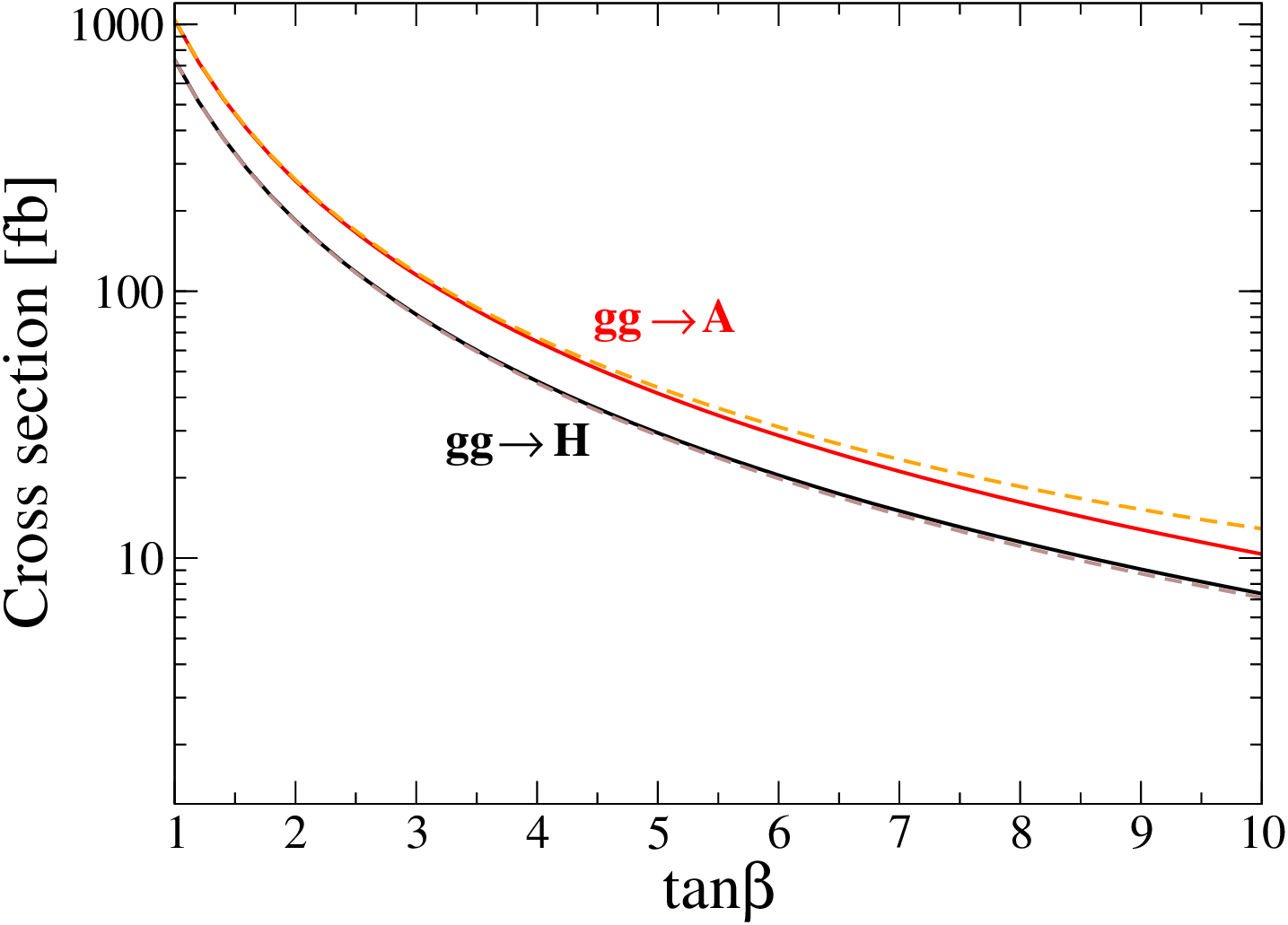}
\caption{Cross section of $gg \to H$ and $gg \to A$ as a function of $\tan\beta$ for the case of $\sin(\beta-\alpha)=1$ and $m_A=m_H=750$ GeV. 
The solid (dashed) curves show the cases with Type-I and Type-X (Type-II and Type-Y) Yukawa interaction. }
\label{fig1}
\end{center}
\end{figure}

Now, we are ready to calculate the gluon-fusion production cross sections of $H$ and $A$. 
In Fig.~\ref{fig1}, we show these as a function of $\tan\beta$ in the case of $\sin(\beta-\alpha)=1$ and 
$m_H^{}=m_A^{}=750$ GeV in all the types of Yukawa interactions.
The cross section is monotonically reduced as $\tan\beta$ increases, 
because the top loop contribution is suppressed due to the $\cot\beta$ factor in the $Ht\bar{t}$ and $At\bar{t}$ couplings. 
The dependence upon the type of Yukawa interaction is not so important in this region of $\tan\beta$. 
In the case of  $\tan\beta < 1$, the top Yukawa coupling becomes too strong to 
guarantee the validity of a perturbative calculation. Besides, such a parameter region is 
strongly constrained from the $B$ physics measurements such as 
$B^0$-$\bar{B}^0$ mixing~\cite{Stal,Watanabe}. 
We thus do not consider the region with $\tan\beta < 1$ in this paper. 

Next, let us discuss the decay rates of the neutral Higgs bosons $H$ and $A$ into the diphoton mode. 
They are given by 
\begin{align}
&\Gamma({\cal H}\to\gamma \gamma)=
\frac{\sqrt{2}G_F\alpha_{\text{em}}^2m_{\cal H}^3}{256\pi^3 }\times \notag\\
& \left|\xi_{\cal H}^V F_1^{\cal H}(m_{W}) + \sum_f Q_f^2N_c^f\xi_{\cal H}^f F_{1/2}^{\cal H}(m_{f})
 -\frac{\lambda_{{\cal H} H^+H^-}}{v}F_0^{\cal H}(m_{H^\pm})
-Q_\chi^2\frac{\lambda_{{\cal H} \chi^\dagger \chi}}{v}F_0^{\cal H}(m_{\chi^\pm})
 \right|^2, \label{Hgg}   \\
&\Gamma(A\to \gamma\gamma)=\frac{\sqrt{2}G_F\alpha_{\text{em}}^2m_A^3}{256\pi^3}\left|\sum_f N_c^f\xi_f Q_f^2 F_{1/2}^A(m_{f})\right|^2, \label{Agg}
\end{align}
where $N_c^f$ is the color factor, $Q_f$ is the electric charge for a fermion $f$ and  
$Q_\chi^2$ represents the squared sum of the electric charges of the charged component field of $\chi$ defined as
\begin{align}
Q_\chi^2 = T^2+(T-1)^2+\cdots +1^2.  \label{Qchi}
\end{align}
The $\xi_{\cal H}^V$ factor denotes the ${\cal H}VV$ ($V=W,~Z$) couplings divided by the 
corresponding $h_{\text{SM}}VV$ couplings, namely, 
$\xi_{h}^V=\sin(\beta-\alpha)$ and $\xi_{H}^V = \cos(\beta-\alpha)$. 
For the charged scalar loops, the scalar trilinear couplings $\lambda_{{\cal H}H^+H^-}$ and $\lambda_{{\cal H}\chi^\dagger \chi}$ 
are defined as the coefficient of the scalar trilinear vertices in the Lagrangian. 
They are expressed as\footnote{Here we use the short-hand notations $s_\beta=\sin\beta$,
$c_\beta=\cos\beta$, $s_{\beta-\alpha}=\sin(\beta-\alpha)$ and $c_{\beta-\alpha}=\cos(\beta-\alpha)$.}
\begin{align}
\lambda_{hH^+H^- }&=\frac{1}{v}\left[(2M^2-2m_{H^\pm}^2-m_h^2)s_{\beta-\alpha}+2(M^2-m_h^2)\cot2\beta c_{\beta-\alpha}  \right],\\
\lambda_{HH^+H^-}&=
-\frac{1}{v}\Big[2(M^2-m_H^2)\cot2\beta s_{\beta-\alpha}+(2m_{H^\pm}^2+m_H^2-2M^2)c_{\beta-\alpha}\Big],\\
\lambda_{h \chi^\dagger \chi} &= -2v[s_{\beta-\alpha}(\rho_1c_\beta^2 + \rho_2 s_\beta^2) - c_{\beta-\alpha}s_\beta c_\beta(\rho_1-\rho_2)], \label{lamh}\\
\lambda_{H \chi^\dagger \chi} &= -2v[c_{\beta-\alpha}(\rho_1c_\beta^2 + \rho_2 s_\beta^2) + s_{\beta-\alpha}s_\beta c_\beta(\rho_1-\rho_2)], 
\end{align}
where $M^2$ denotes the soft-breaking scale of the $Z_2$ symmetry defined by $M^2 = \mu_3^2/(\sin\beta \cos\beta)$~\cite{KOSY}. 
The loop functions in Eq.~(\ref{Hgg}) are given by 
\begin{align}
F_0^{\cal H}(m_{\phi^\pm})  & = \frac{2v^2}{m_{\cal H}^2}[1+2m_{\phi^\pm}^2C_0(0,0,m_{\cal H}^2,m_{\phi^\pm},m_{\phi^\pm},m_{\phi^\pm})],\\
F_1^{\cal H}({m_W})  & = \frac{2m_W^2}{m_{\cal H}^2}\left[6+\frac{m_{\cal H}^2}{m_W^2}+(12m_W^2-6m_{\cal H}^2)C_0(0,0,m_{\cal H}^2,m_W,m_W,m_W)\right]. 
\end{align}

Notice here that the charged scalar loop contributions to the decay rate 
only appear for the case of CP-even Higgs bosons, so that 
the decay rate of $A$ into  diphoton is not enhanced by charged scalar loops. 
Typically, the branching fraction of $A\to \gamma\gamma$ is of ${\cal O}(10^{-5})$ when $\tan\beta\sim 1$ in all  types of Yukawa interactions. 
By looking at Fig.~\ref{fig1}, the production cross section of $gg\to A$ process is about 1 pb at $\tan\beta\sim 1$, so that we obtain 
that $\sigma(gg\to A \to \gamma\gamma) \simeq \sigma(gg\to A)\times \text{BR}(A\to \gamma\gamma)\simeq 0.01$ fb. 
This value is  ${\cal O}(10^3)$ times smaller than the required cross section to explain the diphoton excess. 
For this reason, the CP-odd Higgs boson contribution does not help to explain the diphoton excess, and in the following we concentrate on the 
$H$ contribution, i.e., $gg \to H \to \gamma\gamma$. 

\begin{figure}[t]
\begin{center}
\includegraphics[width=75mm]{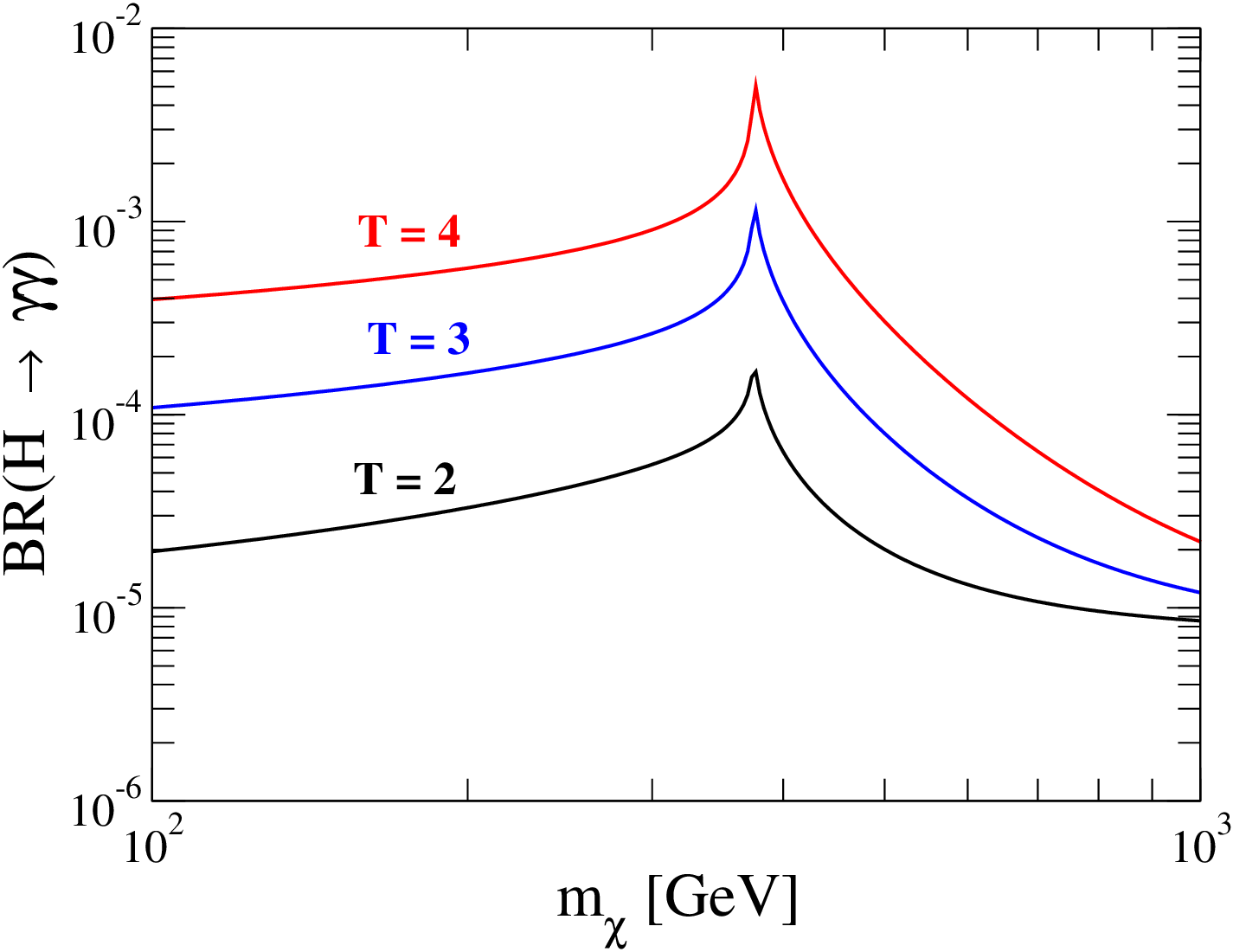} \hspace{5mm}
\includegraphics[width=75mm]{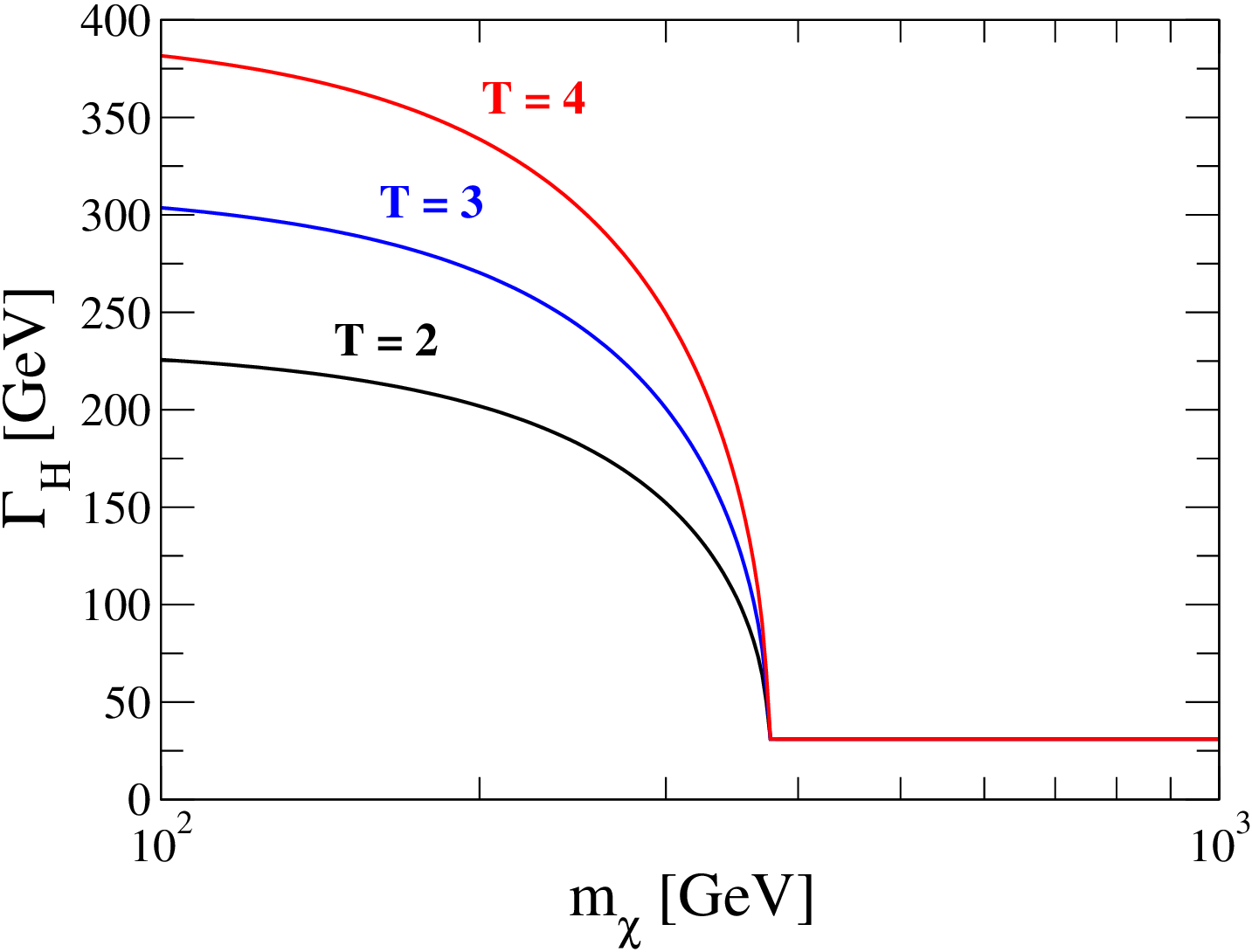}
\caption{(Left) branching fraction of 
$H\to \gamma\gamma$ as a function of $m_\chi$ in the case of 
$\tan\beta=1$, $\sin(\beta-\alpha)=1$ and $m_A=m_{H^\pm}=m_H=M=750$ GeV. 
We fix $\rho_1=-2\sqrt{\pi}$ and  $\rho_2=-\rho_1$.  
(Right) The total width of $H$ as a function of $m_\chi$ for the same parameter set. 
In both frames, the black, blue and red curves  show the cases with $T=2$, 3 and 4, respectively. }
\label{fig2}
\end{center}
\end{figure}

We now show numerical results for the decay rates of the $H$ state, see Fig.~\ref{fig2}. 
In order to obtain the maximal value of the gluon-fusion cross section, we take $\tan\beta =1$. 
In that case, the dependence upon the type of  Yukawa interaction only appears 
in the sign of the bottom quark and tau lepton Yukawa couplings $Hb\bar{b}$ and $H\tau^+\tau^-$ (see Tab.~\ref{xif}), respectively, and 
such a difference can be safely neglected. 
We thus do not specify the type of Yukawa interaction in the following analysis any further. 
In addition, we take the SM-like limit $\sin(\beta-\alpha)=1$ to keep the SM-like Higgs boson $h$ coupling to 
the gauge bosons $hVV$ and fermions $hf\bar{f}$ to be the same values as those in the SM itself. 
Such a situation is favored by the LHC Run-I data~\cite{Run1-ATLAS,Run1-CMS}. 
Therefore, our benchmark scenario to explain the diphoton excess is defined by the following parameter choices:
\begin{align}
m_H^{}=m_A^{}=m_{H^\pm}^{}=M=750~\text{GeV},\quad
\tan\beta = 1,\quad \sin(\beta-\alpha) = 1\quad \text{and}\quad \rho_2 =-\rho_1.  \label{bm}
\end{align}
In the above configuration, we obtain $\lambda_{h\chi^\dagger \chi}=0$ but $\lambda_{H\chi^\dagger \chi}\neq 0$, 
so that the $\chi$ loop effect to the diphoton decay rate only appears in $H$. 
We thus can realize the scenario where 
the branching fraction of $h\to \gamma\gamma$ is kept to be the SM value, while that of 
$H\to \gamma\gamma$ gets a significant contribution from the $\chi$ loop. 
We note that the $H^\pm$ and $W$ boson loop effects to the $H\to \gamma\gamma$ decay vanish in this configuration because $\lambda_{HH^+H^-}=\xi_H^V=0$. 

Before we move on the numerical evaluation for the branching fraction, we briefly mention about constraints on the parameter space from theoretical and experimental sources. 
First, there are bounds from the perturbative unitarity and the vacuum stability which are given by theoretical requirements. 
The former and latter constraints have been discussed in Ref.~\cite{PU} and in Ref.~\cite{VS} in 2HDMs, respectively. 
In the setup given in Eq.~(\ref{bm}), we obtain $\lambda_1=\lambda_2=\lambda_3 = m_h^2/v^2 \simeq 0.26$ and $\lambda_4=\lambda_5=0$. 
This satisfy both the unitarity and vacuum stability constraint in the 2HDM. 
Of course, we cannot simply apply these constraints to our model because of the introduction of the $\chi$ field, i.e., 
we need to take into account the additional dimensionless parameters $\rho_1$, $\rho_2$, $\lambda_\chi$ and $\lambda_\chi'$ in the potential. 
In this paper, we simply argue the perturbativity of these parameters, where the magnitude of these couplings do not exceed a certain value such as 
$2\sqrt{\pi}$ or $4\pi$.

As the experimental constraints, we first consider the electroweak precision valuables such as $S$, $T$ and $U$ parameters~\cite{STU}.  
The additional contributions to $S$ and $T$ parameters are exactly zero under the setup in Eq.~(\ref{bm}) at the one-loop level. 
Besides, if we neglect the small mass difference between $W$ and $Z$ bosons, the new contribution to the $U$ parameter is also exactly canceled. 
Regarding to the flavour constraint, we can apply the same bound in the 2HDM to our model up to the leading order calculation
because $\chi$ does not directly couple to the SM fermions. 
According to the recent study for the flavour constraints in the 2HDM~\cite{Watanabe}, our benchmark point is allowed. 
In Ref.~\cite{singlet-0}, the 95\% CL upper limit on the cross section for $pp \to \phi^0 \to XY$ 
($\phi^0$ being a new neutral  scalar boson with the mass of 750 GeV, and $X$ and $Y$ being the SM particles) 
is shown at the LHC with the collision energy of 8 TeV, where the gluon fusion production cross section at 8 TeV 
is about 5 times smaller than that at 13 TeV.  
We check that 
in our benchmark scenario, all the predictions of the cross section are below the upper bound\footnote{Only for the $\gamma\gamma$ mode, 
our prediction can be compatible to the upper limit on the cross section 
$\sim 1.5$ fb~\cite{singlet-0} in the case where the cross section of the diphoton excess, i.e., $\sim 5$-10 fb at 13 TeV can be explained. } given in \cite{singlet-0}.

In Fig.~\ref{fig2}, 
we show the $m_\chi$ dependence of the branching fraction of $H\to \gamma\gamma$ (left panel) and the total width $\Gamma_H$ of $H$  (right panel). 
We can see in the left figure that the branching fraction becomes the maximum value at around $m_{\chi}\simeq m_H^{}/2\simeq 375$ GeV, because of the threshold effect 
of the $\chi$ loop contribution. 
We note that the $H^\pm$ loop contribution vanishes in the benchmark point given in Eq.~(\ref{bm}) because of $\lambda_{HH^+H^-}=0$. 
By looking at the right figure, we see the drastic difference in $\Gamma_H$ between the cases of $m_\chi < m_H^{}/2$ and $m_\chi > m_H^{}/2$. 
If we take $m_\chi < m_H^{}/2$, the tree level decay mode $H\to \chi^\dagger \chi$ opens, 
where $\chi^\dagger \chi$ includes all the possible combination of the component fields in $\chi$ 
($\chi^\dagger \chi = \chi^0\chi^0,~\chi^+\chi^-, \dots$). 
Conversely, in the case of $m_\chi > m_H^{}/2$, 
the tree level decay is kinematically forbidden and $\Gamma_H$ does not depend strongly upon either the isospin $T$ or the mass $m_\chi$ and is about 30 GeV, indeed a value compatible with the one fitted to the 750 GeV excess.

\begin{figure}[t]
\begin{center}
\includegraphics[width=90mm]{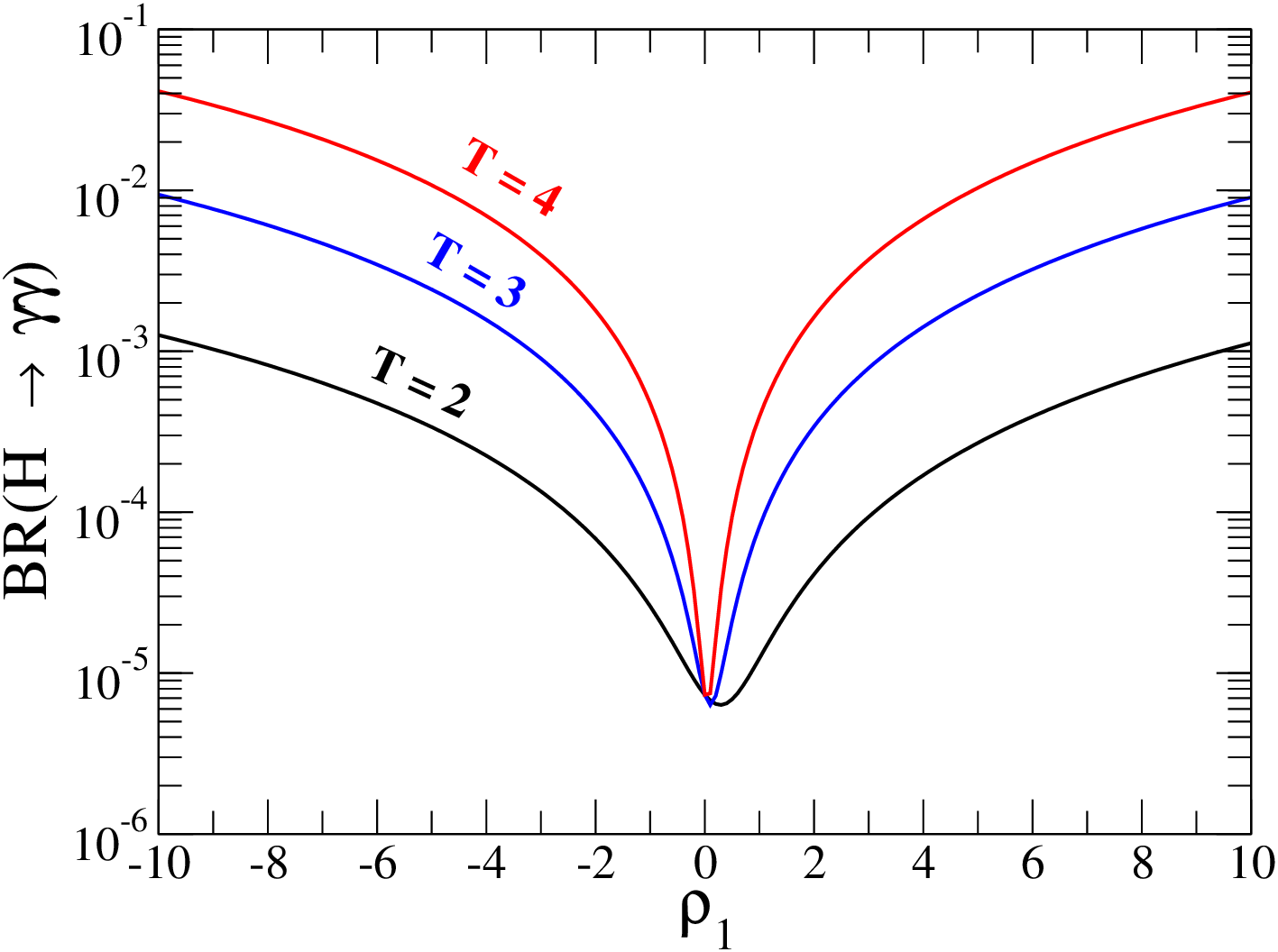}
\end{center}
\caption{Branching fraction of $H\to \gamma\gamma$ as a function of $\rho_1\,(=-\rho_2)$ 
in the case of $\tan\beta=1$, $\sin(\beta-\alpha)=1$ and $m_A=m_{H^\pm}=m_H=M=750$ GeV. 
The black, blue and red curves  show the cases with $T=2$, 3 and 4, respectively. 
The mass of $\chi$ is fixed to be 375 GeV. }
\label{fig3}
\end{figure}

In Fig.~\ref{fig3}, we show the $\rho_1$ dependence of the branching fraction of the $H\to \gamma\gamma$ mode. 
We take $m_\chi=375$ GeV to extract the maximal value of the branching fraction for each fixed value of $\rho_1$. 
We note that a slightly larger value of the branching fraction is obtained when we take negative values of $\rho_1$ as compared to the case 
of positive values with the same magnitude.
The reason is that the contributions between the top loop and the $\chi$ loop becomes constructive (destructive) when
we take $\rho_1<0$ ($\rho_1 >0$).
We find that, in order to obtain the branching fraction to be the order of $10^{-2}$,  
we need $|\rho_1|\gtrsim 10~(4)$ for the case of $T=3~(4)$. 

\begin{figure}[t]
\begin{center}
\includegraphics[width=50mm]{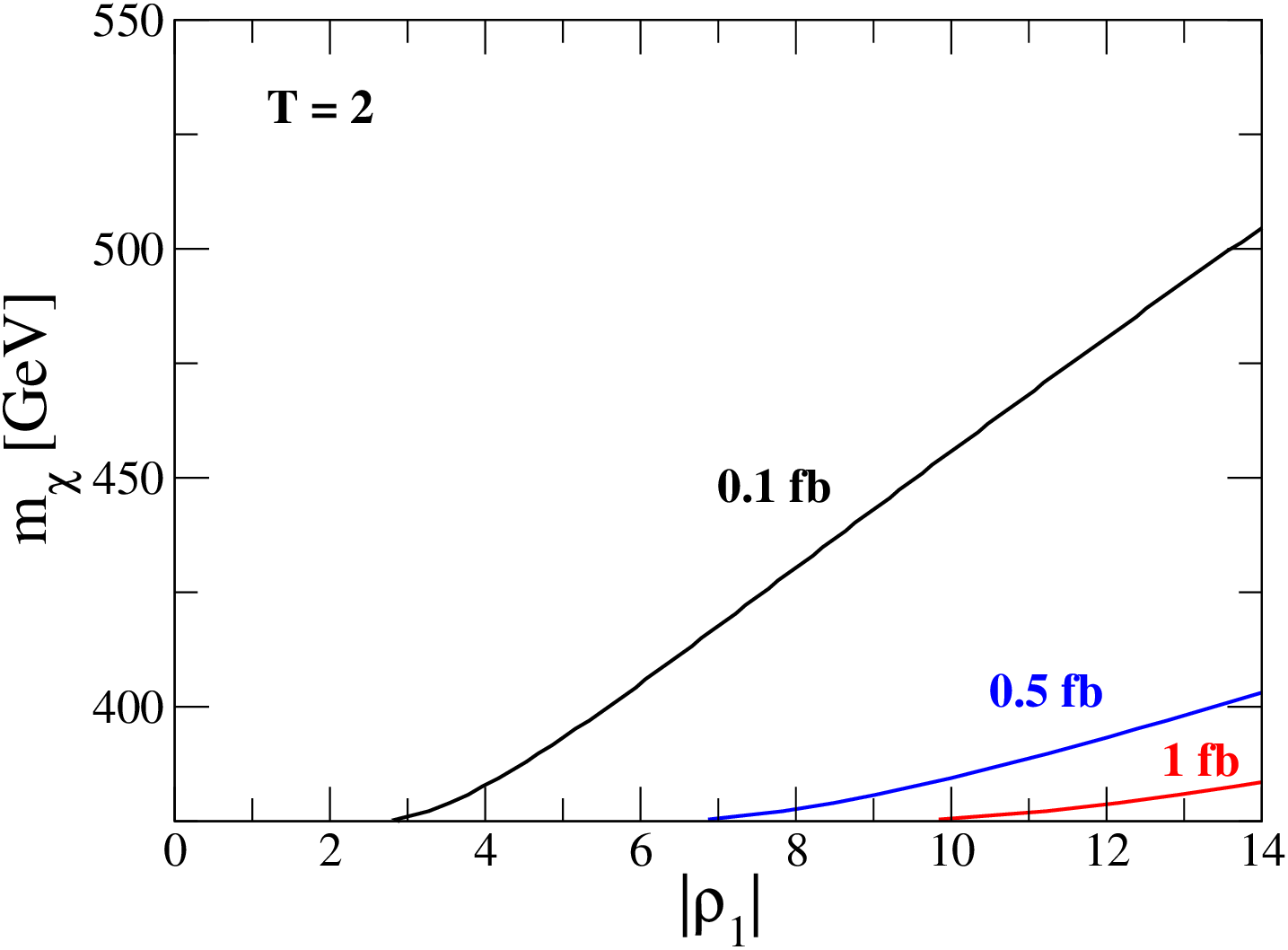}\hspace{3mm}
\includegraphics[width=50mm]{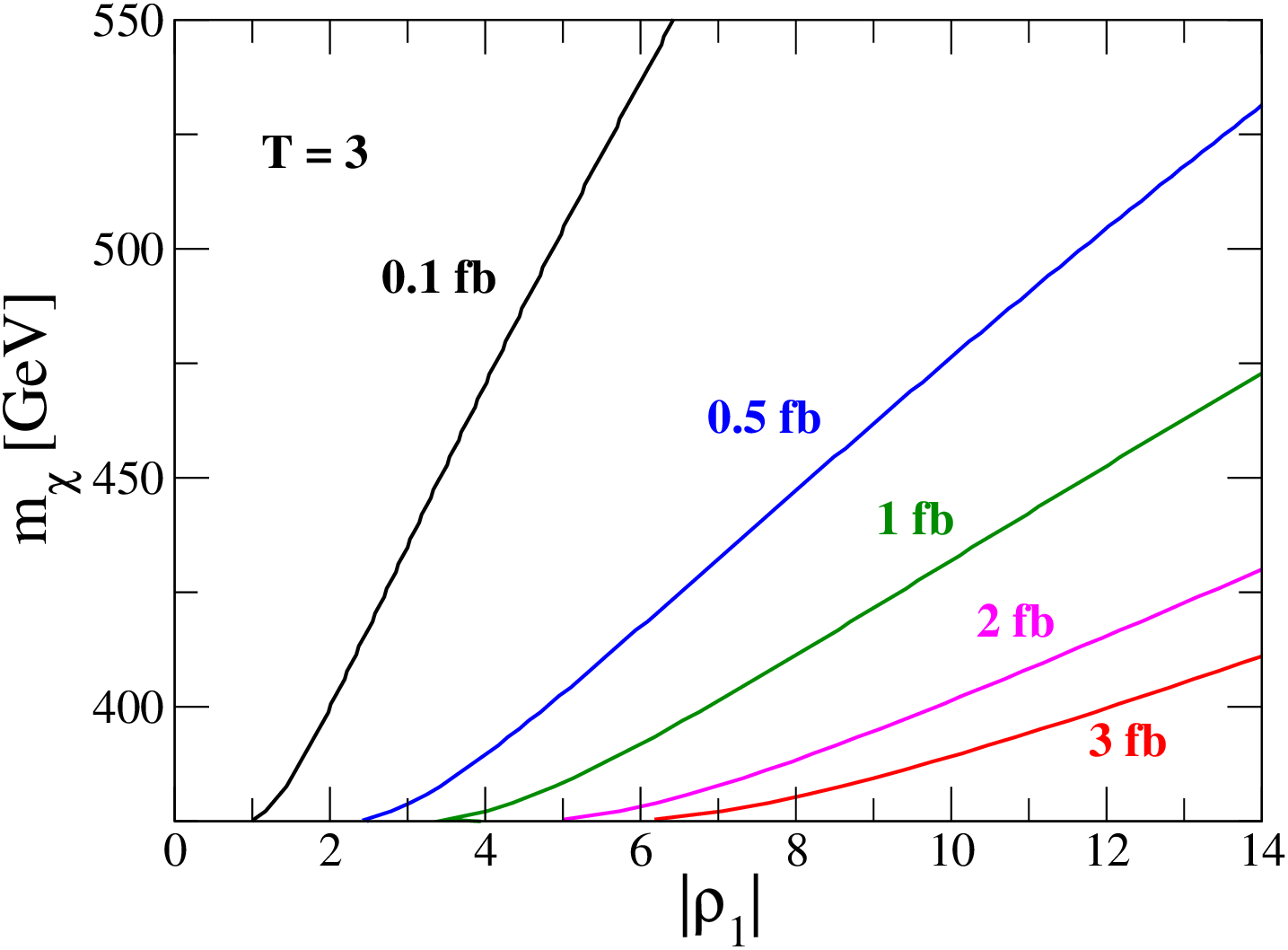}\hspace{3mm}
\includegraphics[width=50mm]{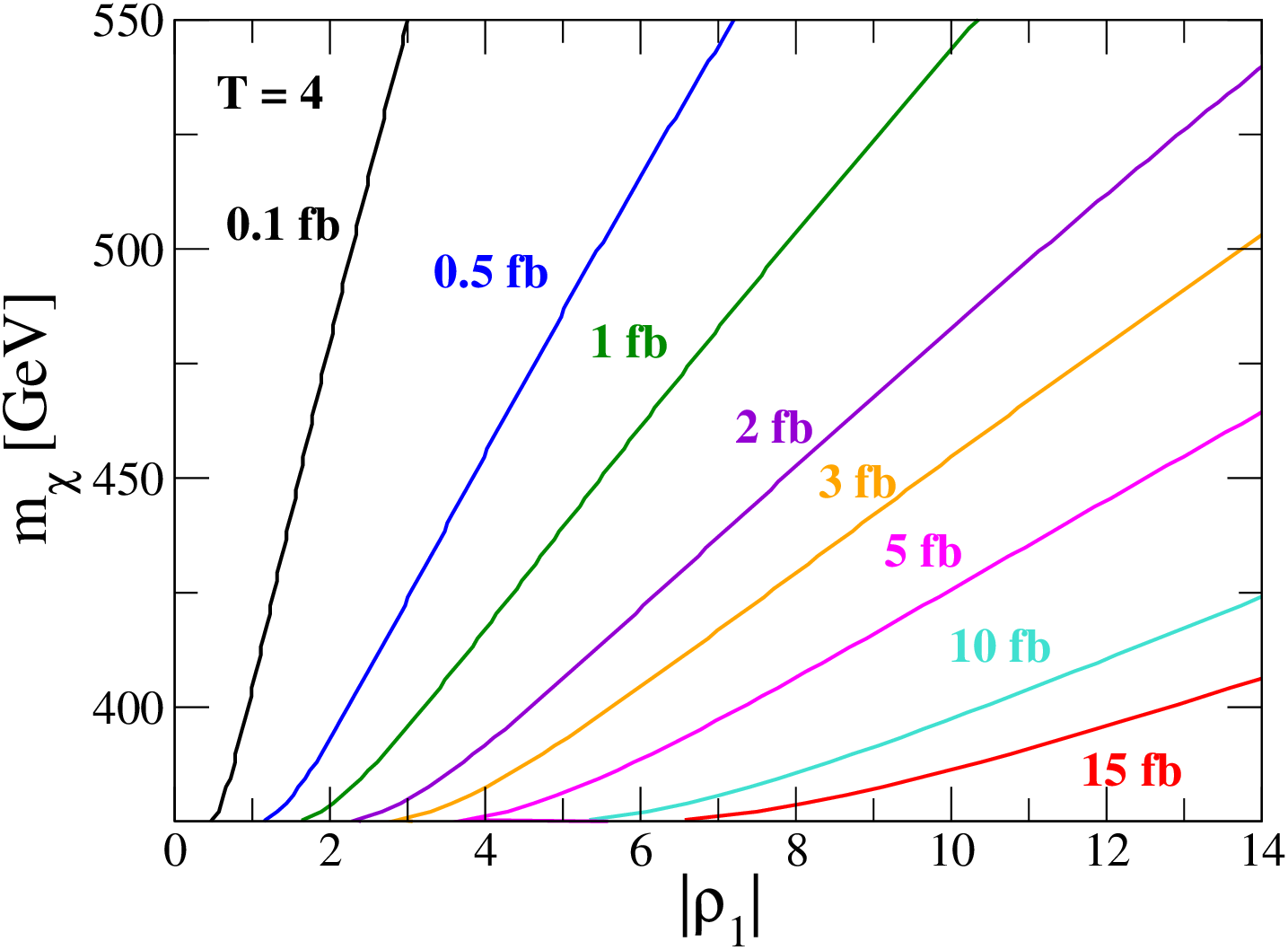}
\caption{Contour plots for the cross section of $gg \to H \to \gamma\gamma$ on the $|\rho_1|$-$m_\chi$ plane with $\rho_1<0$ and $\rho_2 =-\rho_1$. 
We take $\tan\beta=1$, $\sin(\beta-\alpha)=1$ and $m_A=m_{H^\pm}=m_H=M=750$ GeV. 
The left, center and right panels show the cases with $T=2$, 3 and 4, respectively. }
\label{fig4}
\end{center}
\end{figure}

In Fig.~\ref{fig4}, we show the contour plots for the cross section of the $gg \to H \to \gamma\gamma$ process, where we use the narrow
width approximation. 
The results for $T=2$, 3 and 4 are shown in the left, center and right panel, respectively. 
We find that in the case of $\rho_1 = -2\sqrt{\pi}$ and $m_\chi=375$ GeV, 
the cross section is given to be about 0.1 fb, 1 fb and 3 fb for $T=2$, 3 and 4, respectively. 

Finally, we discuss the cutoff scale $\Lambda$ in our model, wherein
running coupling constants become infinity, namely, we have a Landau pole at $\Lambda$ and 
our model should be replaced by a fundamental theory which is expected to describe physics from $\Lambda$ to the Planck scale. 
In our model, we expect the Landau pole to appear well below the Planck scale, because we introduced 
rather large coupling constants, $\rho_1$ and $\rho_2$, plus a high isospin scalar multiplet $\chi$.  
In order to know the cutoff scale, 
we calculate the running coupling constants by solving the RGEs at the one-loop level.  
In Appendix, we give the one-loop beta functions for all the coupling constants. 
We take the initial scale to be $m_H^{}=750$ GeV and the set of initial values of coupling constants as follows
\begin{align}
&\lambda_1(m_H^{}) = \lambda_2(m_H^{}) = \lambda_3(m_H^{}) = \frac{m_h^2}{v^2},~ 
\lambda_4(m_H^{}) = \lambda_5(m_H^{}) = 0, ~y_t(m_H^{}) = \frac{\sqrt{2}m_t}{v}, \notag\\
&\rho_2(m_H^{})=-\rho_1(m_{Z}^{}),~\lambda_\chi(m_H^{}) = \lambda_\chi'(m_H^{}) = 0, \label{initial}
\end{align}
with $m_h=125$ GeV and  $m_t=173.21$ GeV. 
Instead of the extraction of the scale $\lambda_i(\Lambda)=\infty$ ($\lambda_i$ are the coupling constants in our model), 
we define the cutoff scale\footnote{The choice of this critical value, taking $4\pi$, $8\pi$ and so on, 
is not so important, because once one of the coupling constants exceeds ${\cal O}(10)$, then such a coupling quite rapidly blows up.  } to be the scale giving $\lambda_i(\Lambda)=4\pi$. 

\begin{figure}[t]
\begin{center}
\includegraphics[width=70mm]{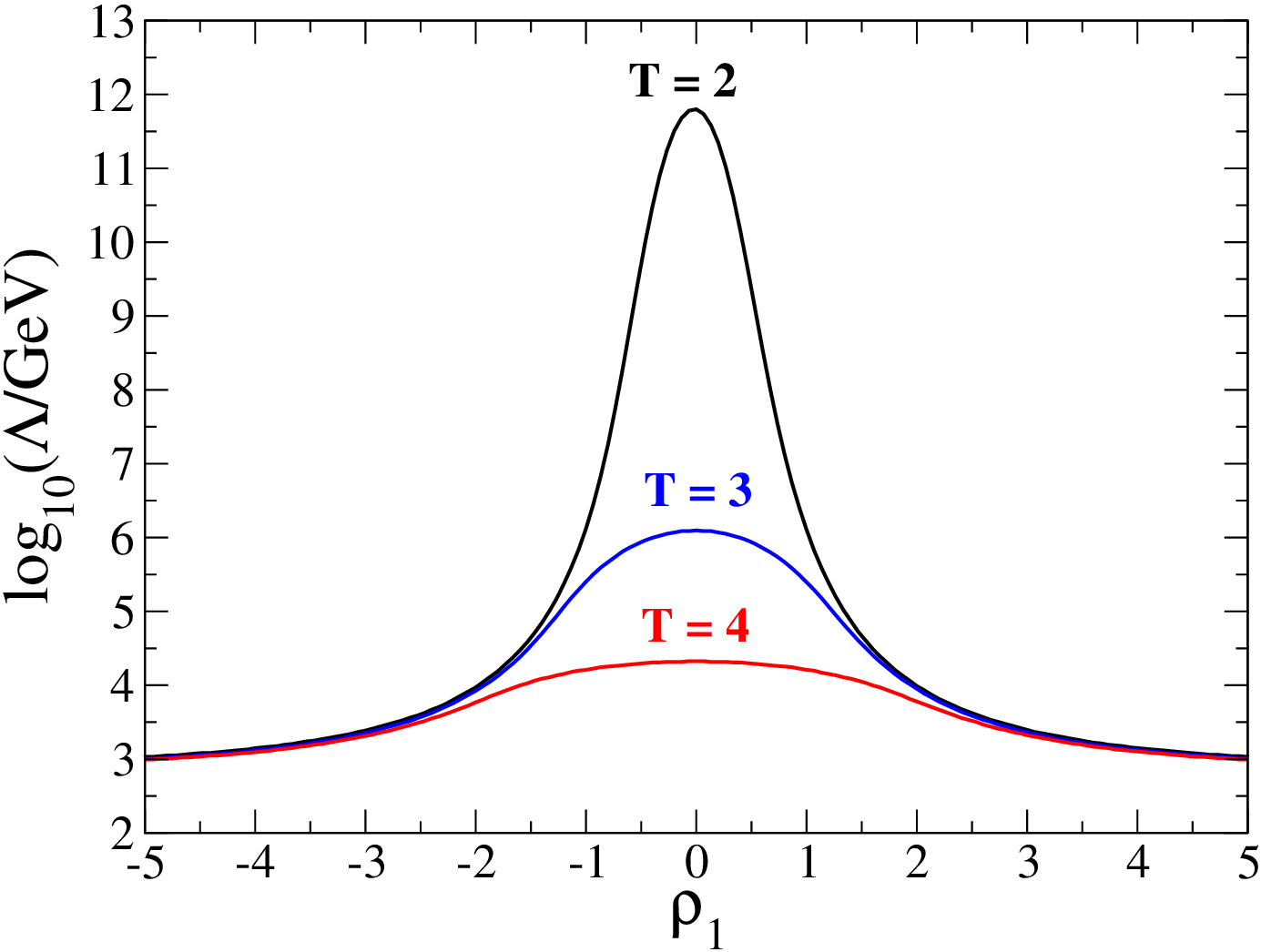}
\end{center}
\caption{The cutoff scale $\Lambda$ as a function of $\rho_1$ for the case of $T=2$ (black), $T=3$ (blue) and $T=4$ (red).
We use the set of initial values of the running coupling constants given in Eq.~(\ref{initial}). }
\label{cutoff}
\end{figure}

In Fig.~\ref{cutoff}, we show the cutoff scale $\Lambda$ as a function of $\rho_1$ for each case of $T=2$, 3 and 4. 
We see that the model with a larger isospin multiplet has $\Lambda$ at lower energy scale as it is expected.
However, once $|\rho_1| \gtrsim 2$ is taken, $\Lambda$ is given to be below the order 10 TeV scale in all the cases. 
Therefore, if we require $\Lambda>1$ TeV, then the maximally allowed cross section of the diphoton process is 
about less than 0.1 fb, 0.5 fb and 1 fb in the case of $T=2$, 3 and 4, respectively. 

Before closing this section, we would like to briefly comment on the possibility of one-loop induced
$H\to VV$ ($V=W,Z$) decays. 
Although in the above analysis we take $\sin(\beta-\alpha)=1$ in order to eliminate the tree level $H\to VV$ decays, these
could appear at the one-loop level, where the top quark runs in both the $H\to W^+W^-$ and  $H\to ZZ$ diagrams, while 
$\chi$ does so only in the $H\to W^+W^-$ diagram (there are no $\chi^\dagger \chi Z$ and $\chi^\dagger \chi ZZ$ couplings due to the real scalar nature of $\chi$ ). 
Since $\chi$ has a large isospin charge and a large coupling to the $H$ state, we expect  the $\chi$ loop to give a significant 
contribution to the $H\to W^+W^-$ decay, which can reach the upper limit on the cross section of $pp\to H\to W^+W^-$. 
The upper limit was derived to be 40 fb with 95\% CL from the LHC Run-I data~\cite{singlet-0}. 
The one-loop induced decay rate of $H\to W^+W^-$ is expressed as 
\begin{align}
\Gamma(H \to W^+W^-)& = \left(\frac{g^2}{16\pi^2}\right)^2\frac{1}{16\pi m_H^{}}|c_{\text{eff}}^{HWW}|^2\left(3-\frac{m_H^2}{m_W^2}+\frac{m_H^4}{4m_W^4}\right)\sqrt{1-\frac{4m_W^2}{m_H^2}}\notag\\
&\simeq (0.19~\text{GeV})\times \frac{|c_{\text{eff}}^{HWW}|^2}{m_H^2}, 
\end{align}
where $c_{\text{eff}}^{HWW}$ is the dimensionful one-loop induced $HW^+W^-$ coupling factorized by $g^2/(16\pi^2)$. 
Furthermore, the cross section of the $gg\to H$ process is about 160 fb~\cite{SM-cross} at 8 TeV, 
which sets a 95\% CL upper bound on the branching fraction of $H\to W^+W^-$ mode at about 0.25. 
Using the total width of the $H$ state to be 30 GeV, we obtain the upper bound on $|c_{\text{eff}}^{HWW}|$ to be about 4.7 TeV. 
Suppose that the effective coupling $c_{\text{eff}}^{HWW}$ is naively given by $\sum_{Q=0}^{T} [T(T+1)-Q]^2\lambda_{H\chi^{Q*}\chi^Q}v^2C_0(m_W^2,m_W^2,m_H^2;m_\chi,m_\chi,m_\chi)$
then we get $|c_{\text{eff}}^{HWW}|$ to be less than 2.9 TeV, 8.7 TeV and 19 TeV for the case of $T=2$, 3 and 4, with $\rho_1=1~(\rho_2 = -\rho_1)$. 
Therefore, our prediction of the branching fraction of $H\to W^+W^-$ could be comparable or even above the upper bound. 
However, in order to get a more precise result, we would need to perform the renormalization prescription of the $HW^+W^-$ vertex as well as take into account 
the destructive top loop effect, which is beyond the scope of the current paper. 

\section{Conclusions}

We have discussed the extension of the softly-broken $Z_2$ symmetric 2HDMs, where the scalar sector is composed of two active complex doublets 
and a real inert scalar multiplet $\chi$ with an isospin $T(=2,3,4)$ whose lightest neutral component could be 
a DM candidate. 
In this model, we have investigated the possibility to explain the diphoton excess at around 750 GeV recently observed by the LHC experiments with 13 TeV energy. 
In our model, the additional active neutral Higgs boson $H$ and $A$ contribute to the diphoton process via $gg\to H/A \to \gamma\gamma$. 
In order to explain this excess, we need an ${\cal O}(10^3)$ enhancement 
in the branching fractions of $H/A \to \gamma\gamma$ as compared to the standard 2HDMs. 
Such an enhancement can be realized by the loop effects induced by charged inert particles.  
We have shown that the CP-odd Higgs boson $A$ contribution does not help to explain the excess because the branching fraction of $A\to \gamma\gamma$ 
cannot be enhanced by the $\chi$ loop diagrams. 
In contrast, the branching fraction of $H\to \gamma\gamma$ can be significantly enhanced by such a $\chi$ loop effect. 
Upon using the narrow width approximation, 
we have found that, if we take the coupling $\rho_1$ to be $-2\sqrt{\pi}$, 
the cross section $gg \to H\to \gamma\gamma$ can be about 0.1 fb, 1 fb and 3 fb in the case of $T=2$, 3 and 4, respectively. 
When we allow rather strong couplings such as $\rho_1 = -6$, we then obtain 
a cross section of about 0.5 fb , 3 fb and 10 fb in the case of $T=2$, 3 and 4, respectively.
We have checked where the cutoff scale $\Lambda$ appears in our model. 
By solving the RGEs at the one-loop level, we obtained 
$\Lambda$ to be about $10^{12}$, $10^6$ and $10^{4.5}$ GeV for $T=2$, 3 and 4 when $\rho_1\simeq 0$.
This scale becomes smaller than ${\cal O}(10)$ TeV when we take $|\rho_1|\gtrsim 2$ for $T=2$, 3 and 4, and, 
if we require $\Lambda>10$ TeV, then the maximally allowed cross section of the diphoton process is 
less than about 0.1 fb, 0.5 fb and 1 fb in the case of $T=2$, 3 and 4, respectively.

\vspace*{4mm}
\noindent
\section*{Acknowledgments}
\noindent
 S. M. is supported in part through the NExT Institute. 
 K.~Y. is fully supported by a JSPS postdoctoral fellowships for research abroad. Both authors acknowledge discussions with
S.F.  King.

\noindent
\section*{Note added}
\noindent
After this paper was completed,
Ref.~\cite{Han} appeared in which the diphoton excess was discussed in a model 
with two Higgs doublet fields and a real inert septet field. 

\begin{appendix}
\section{One-loop beta functions}

In this Appendix, we give the analytic formulae for the 
beta functions at the one-loop level which are used for the RGE analysis given in Sec.~III. 
The beta functions are defined as 
\begin{align}
\beta(\lambda_i) \equiv \frac{d}{d\ln \mu}\lambda_i, 
\end{align}
where $\mu$ is an energy scale. 
For the Yukawa couplings, we only keep the contribution of the top Yukawa coupling $y_t$.  

For the gauge couplings, 
the beta functions for the $SU(3)_c$ $(g_s)$, $SU(2)_L$ $(g)$ and $U(1)_Y$ $(g_Y^{})$ coupling are given by
\begin{align}
\beta (g_s) &= -\frac{1}{16\pi^2}7g_s^3, \\
\beta (g) &= \frac{1}{16\pi^2}\left[-\frac{19}{6}+\frac{2}{3}(1^2+2^2+\cdots+T^2)\right]g^3, \\
\beta (g_Y) &= \frac{1}{16\pi^2}7g_Y^3,
\end{align}
where $\beta(g_s)$ is the same form as the SM one. 

The beta function for $y_t$ is given by
\begin{align}
\beta (y_t) &= \frac{1}{16\pi^2}\left[ 9y_t^3-\left(8g_s^2+\frac{9}{4}g^2+\frac{17}{12}g_Y^2\right)y_t \right]. 
\end{align}

Finally, we give the beta functions for all the scalar quartic couplings as follows
\begin{align}
\beta(\lambda_1)&=\frac{1}{16\pi^2}\Big[12\lambda_1^2+4\lambda_3^2+2\lambda_4^2+4\lambda_3\lambda_4+2\lambda_5^2 +4\rho_1^2(1+2T)\notag\\
&\quad\quad\quad\quad+\frac{9}{4}g^4+\frac{3}{4}g_Y^4+\frac{3}{2}g_Y^2g^{2}-3\lambda_1 (3g^2+ g_Y^2) \Big], \\
\beta(\lambda_2)&=\frac{1}{16\pi^2}\Big[12\lambda_2^2+4\lambda_3^2+2\lambda_4^2+4\lambda_3\lambda_4+2\lambda_5^2+4\rho_2^2(1+2T) -12y_t^4\notag\\
&\quad\quad\quad\quad
+\frac{9}{4}g^4+\frac{3}{4}g_Y^4+\frac{3}{2}g_Y^2g^{2}-3\lambda_2 (3g^2+g_Y^2-4y_t^2) \Big], \\
\beta(\lambda_3)&=\frac{1}{16\pi^2}\Big[2(\lambda_1+\lambda_2)(3\lambda_3+\lambda_4)+4\lambda_3^2+2\lambda_4^2+2\lambda_5^2
+4\rho_1\rho_2(1+2T)\notag\\
&\quad\quad\quad\quad+\frac{9}{4}g^4+\frac{3}{4}g_Y^4-\frac{3}{2}g_Y^2g^2
-3\lambda_3(3g^2+g_Y^2-2y_t^2)\Big], \\
\beta(\lambda_4)&=\frac{1}{16\pi^2}\Big[2\lambda_4(\lambda_1+\lambda_2+4\lambda_3+2\lambda_4)+8\lambda_5^2+3g^2g_Y^2
-3\lambda_4\big(3g^2+g_Y^2-2y_t^2\big)\Big], \\
\beta(\lambda_5)&=\frac{1}{16\pi^2}\Big[2\lambda_5(\lambda_1+\lambda_2+4\lambda_3+6\lambda_4)-3\lambda_5(3g^2+g_Y^2-2y_t^2)\Big], \\
\beta(\rho_1)&=\frac{1}{16\pi^2}\Big[\frac{3}{2}T(T+1)g^4+6\lambda_1\rho_1 + 4\lambda_3\rho_2 + 2\lambda_4\rho_2+ 8\rho_1^2 + 8\rho_1\lambda_\chi(3+2T) \notag\\
&\quad\quad\quad\quad  + 4\rho_1\lambda_\chi'\Big(6c_0 + \sum_{i=1,T}c_i\Big) -\rho_1\big(6g^2T(T+1)+\frac{9}{2}g^2+\frac{3}{2}g_Y^2-6y_t^2\big)\Big], \\
\beta(\rho_2)&=\frac{1}{16\pi^2}\Big[\frac{3}{2}T(T+1)g^4+6\lambda_2\rho_2 + 4\lambda_3\rho_1 + 2\lambda_4\rho_1+ 8\rho_2^2+ 8\rho_2\lambda_\chi(3+2T)\notag\\
&\quad\quad\quad\quad  + 4\rho_2\lambda_\chi'\Big(6c_0 + \sum_{i=1,T}c_i\Big)-\rho_2\Big(6g^2T(T+1)+\frac{9}{2}g^2+\frac{3}{2}g_Yxs^2-6y_t^2\Big)\Big], \\
\beta(\lambda_\chi) &=\frac{1}{16\pi^2}\Big[\frac{3}{4}T^2(T+1)^2g^4 + 2(\rho_1^2+\rho_2^2)+ 72(\lambda_\chi+c_0\lambda_\chi')^2+ \sum_{i=1,T}(4\lambda_\chi+c_i \lambda_\chi')^2 \notag\\
&\quad\quad\quad\quad-12g^2T(T+1)(\lambda_\chi + c_0\lambda_\chi')\Big]-c_0\beta(\lambda_\chi'), 
\end{align}
where
\begin{align}
&(c_0,c_1,c_2) = (0,0,0)~\text{for}~~T=2, \\
&(c_0,c_1,c_2,c_3) = (72,288,432,72)~\text{for}~~T=3, \\
&(c_0,c_1,c_2,c_3,c_4) = (200,800,1360,440,160)~\text{for}~~T=4. 
\end{align}
The beta function for $\lambda_\chi'$ is respectively given for the $T=3$ and $T=4$ case as 
\begin{align}
\beta(\lambda_\chi') &=\frac{1}{16\pi^2}\left(6120\lambda_\chi^{\prime 2} + 96 \lambda_\chi\lambda_\chi' +\frac{3}{2}g^4-144 g^2\lambda_\chi'\right),\quad \text{for}~~T=3, \\
\beta(\lambda_\chi') &=\frac{1}{16\pi^2}\left(21992\lambda_\chi^{\prime 2} + 96 \lambda_\chi\lambda_\chi' +\frac{3}{2}g^4-240 g^2\lambda_\chi'\right),\quad \text{for}~~T=4.  
\end{align}
We have checked the consistency of the above formulae with those given in Ref.~\cite{Tsumura}. 

\end{appendix}

\end{document}